\documentstyle[aps,multicol,epsf]{revtex}
\draft
\def\beginwide{
        \end{multicols} \vspace*{-0.5cm} \noindent
        \rule{3.5in}{.1mm}\rule{.1mm}{5mm} \widetext \medskip }
\def\beginwidetop{
        \end{multicols} \vspace*{-0.5cm} \noindent
        \widetext \medskip }
\def\endwide{
        \hspace*{3.35in}~\rule[-5mm]{.1mm}{5mm}\rule{3.5in}{.1mm}
        \begin{multicols}{2} \vspace*{-1.0cm} \noindent }
\def\endwidebottom{
        \begin{multicols}{2} \vspace*{-1.0cm} \noindent }
\newcommand{\beq}{\begin{equation}}
\newcommand{\eeq}{\end{equation}}
\newcommand{\bdis}{\begin{displaymath}}
\newcommand{\edis}{\end{displaymath}}
\newcommand{\bea}{\begin{eqnarray}}
\newcommand{\eea}{\end{eqnarray}}
\newcommand{\barr}{\begin{array}}
\newcommand{\earr}{\end{array}}
\begin{document}

\title{Order Parameter and Scaling Fields in 
Self-Organized Criticality}

\author{Alessandro Vespignani$^{1}$ and  Stefano Zapperi$^{2}$}

\address{$^1$Instituut-Lorentz, University of Leiden, P.O. Box 9506
2300 RA, Leiden, The Netherlands\\
$^2$Center for Polymer Studies and Department of Physics,
Boston University, Boston, MA 02215}

\date{\today}

\maketitle
\begin{abstract}
We present a unified  dynamical mean-field theory for 
stochastic self-organized critical models. We use a single site 
approximation and we include the details of different models 
by using effective parameters and constraints. We identify the order
parameter and the relevant scaling fields in order to
describe the critical behavior in terms 
of usual concepts of non equilibrium lattice models with steady-states. 
We point out the inconsistencies of previous mean-field
approaches, which lead to different predictions.
Numerical simulations confirm the validity of our results
beyond mean-field theory.
\end{abstract}


%
%
%
%

\begin{multicols}{2}
The origin of scaling in nature \cite{bb} has become
in the last years a challenging problem in physics.
Bak, Tang and Wiesenfeld (BTW) \cite{bak2} have proposed self-organized
criticality (SOC) as a unifying theoretical framework to describe
a vast class of driven systems that evolve ``spontaneously''
to a stationary state, characterized by power law distributions
of dissipation events. Despite the insights SOC concepts
have brought to a number of problems, an
agreement on the exact definition of SOC has still not been
reached and the exact meaning of the word ``spontaneous''  is 
quite unclear. Originally, SOC was associated with
the absence of tuning parameters, but
it has been noted \cite{grin,vzl} that
the driving rate acts as a tuning parameter in most, if
not all, SOC models.
This ambiguity has hindered the formulation
of precise relations between SOC and other non equilibrium
critical phenomena \cite{zia,torre}.

In this letter we reformulate SOC in terms of typical
concepts of non equilibrium critical phenomena \cite{torre}
by using the dynamical single site mean-field (MF) theory\cite{dick}. 
We provide a general scheme in which the details of different models are 
included via effective parameters and constraints. We mainly discuss
sandpile models \cite{bak2} with and without dissipation \cite{diss}, 
but the formalism can be directly applied 
to other stochastic SOC models, 
such as the forest-fire model\cite{dro}.
Work is in progress to apply this method also  to systems
driven by an extremal dynamics\cite{extrem}.
We find two independent critical parameters, i.e. relevant scaling fields, 
both with critical value equal to zero, and just in this double 
limit criticality is reached. We study the behavior of the 
order parameter and evaluate critical exponents. 
The results we obtain are in contrast with 
previous MF approaches \cite{bakmf,stellamf}. 
This is due to a subtle inconsistency 
in the way critical parameters have been chosen in previous works.
We show that some  MF exponents are exact also in low dimensional
systems because of conservation laws. 
Our prediction are confirmed by numerical simulations
of two dimensional sandpile models. 

Sandpile models are cellular automata with an integer
(or continuous) variable $z_i$ (energy) defined in a $d-$dimensional
lattice. At each time step
an energy grain is added to a randomly chosen site, until
the energy of a site reaches a threshold $z_c$.
When this happens the site relaxes ($z_i\to z_i -z_c$) 
and  energy is transferred
to the nearest neighbors ($z_j\to z_j +y_j$). For conservative models the
transferred energy equals the energy
lost by the relaxing site ($\sum y_j=z_c$), at least on average.
Usually, the only form of dissipation occurs at
the boundary, from which energy can leave the system.
With these conditions the system reaches a stationary
state characterized by avalanches whose sizes $s$ 
are distributed as a power law $P(s)\sim s^{-\tau}$\cite{bak2,grasma,manna}.

In order to simplify the description of 
these models, we can reduce the number of states each site 
can assume in the following way. We divide sites in {\em critical},
{\em stable} and {\em active} \cite{manna}.
Stable sites are those that do not become active if 
energy is added to them. 
Critical sites become active by the addition of energy. 
Active sites are relaxing and transfer energy, providing
an interaction with other sites, usually the nearest neighbors (n.n.).
In this way we have mapped the system in a three states cellular
automaton (CA) on a $d$-dimensional lattice \cite{torre,jmendes}.
To each site $i$
is associated a variable $s_i$, which can assume three different values.
A complete set $s\equiv\{s_i\}$ of 
lattice variables specifies a configuration of the system. 
The dynamics is characterized by 
the operator $\langle s\mid W \mid s^0\rangle$
which represents the transition rate
from a configuration $s_0$ to
a configuration $s$ in a time step $t$. 
A well established technique to study these systems is the single 
site mean-field approximation \cite{dick}. 
Denoting by $\rho_a,\rho_c$ and $\rho_s$
the average densities of sites in the active, critical and stable states 
respectively, we write the following reaction rate
equations 
\beq
\frac{\partial}{\partial t}\rho_\kappa=F_\kappa(\rho_a,\rho_c,\rho_s)~~~~~
~~~~~~~\kappa=a,c,s.
\label{ssmf}
\eeq
Because the densities must preserve normalization, two 
of the above equations supplemented with the condition 
$\rho_a+\rho_c+\rho_s=1$, are enough to describe completely the system.
The most general way to represent the function $F_\kappa$ is 
through the Taylor series of the average densities:
\beq
F_\kappa= \sum_n f_\kappa^n\rho_n +\sum_{n,\ell}f_\kappa^{n,\ell}
\rho_n\rho_\ell + {\cal O}(\rho_n^3),
\eeq 
where the constant term is set to zero in order to get a stationary 
state. The first order terms are the transition rates generated by
the external driving fields or by spontaneous transitions. 
The second and higher 
order terms characterize  transitions due to the 
interaction between different sites. 
In SOC models, only the active state generates a 
non trivial dynamical evolution, 
while stable or critical sites can change 
their state only because of the external field or the presence of an active 
n.n. site. Since the critical point is identified by $\rho_a = 0$, 
in correspondence with a vanishing external field,
we can neglect second order terms in the density of active sites.
The solutions of the stationary equations
($\frac{\partial}{\partial t}\rho_\kappa=0$)
are function of the effective parameters $f_\kappa^n, f_\kappa^{n,\ell}$,
which depend on the details of the model.
It is expected that the critical behavior is not affected by the 
specific values of the parameters, while
universality classes will depend on constraints 
imposed on the equations, because of symmetries and  conservation laws.

For the seek of clarity, we describe in details the case of 
sandpile models. 
In this class  of systems the only external field is the flow  of 
energy added to the system. We can describe this driving by  
the probability per unit time $h$ that a site will receive 
a grain of energy. The total amount of energy 
added to the system at each time step will be $J_{in}=h L^d$.
The first order terms in $F_a$ are 
the transition rates $a\to s,c$ and vice versa,
independently of nearest neighbor sites:
\beq
f_a^a=-1 ~~~;~~~f_a^s=0 ~~~;~~~f_a^c=h.
\eeq
Here we considered that active sites becomes stable with 
unitary rate, stable sites never becomes active and critical sites 
becomes active because of the external field.
In addition, there is a single interaction 
term that describes the creation
of an active site from a critical site due to the relaxation
of n.n. sites. We can write this term as 
$(g-\epsilon)\rho_c\rho_a$, where $g$ is an effective rate
that depends on the geometry and the energy involved in the 
relaxation process and $\epsilon$ is
the average energy dissipated in each site\cite{vz}. We stress 
that $\epsilon$ is present also for fully conservative systems,
being an effective term due to the boundary dissipation \cite{disboun}.  
Considering all these terms we obtain 
\beq
F_a= -\rho_a +h\rho_c +(g-\epsilon)\rho_c\rho_a + {\cal O}(\rho_a^2). 
\eeq
A similar reasoning yields the functions $F_c$ and $F_s$. 
The effect of  the driving field on stable sites and
the interaction between active and stable sites deserves
a discussion. The corresponding terms are  
contributing to the transition rate $s\to c$. In sandpiles models,
the energy conservation imposes a local constraint in the rate 
equations. Energy is stored in stable sites until they become critical,
but only a fraction $u$ of stable sites receiving an energy grain
contribute to the $s\to c $ process. 
Therefore in this case the reaction rates will be given by the $c\to a$ rates
multiplied by the factor $u$. For instance, a stable site will receive an
energy grain with probability $h$, but only with probability $uh$ it will 
turn critical. The same reasoning holds for the interaction term.

After imposing stationarity,
we get the following dynamical MF equations:
\bea
\nonumber
\rho_a =h\rho_c +(g-\epsilon)\rho_c\rho_a\\
\nonumber
\rho_a = uh\rho_s + u(g-\epsilon)\rho_s\rho_a\\
\rho_a =1 - \rho_s - \rho_c
\label{mf3}
\eea
where $u,g$ are effective parameters which depend upon the particular model,
$\epsilon$ represents the 
dissipation and $h$ is the driving field. We expect $g$
to be an independent parameter of the model, 
while $u$ has to be obtained self-consistently, because it is fixed once the 
dynamical rules of the CA are given.

After some algebra from Eqs.~(\ref{mf3}) 
we obtain a closed equation for $\rho_a$ 
\beq
u(g-\epsilon)\rho_a^2 +(1+u(1+h-g+\epsilon))\rho_a -uh=0.
\eeq
We can expand $\rho_a(h)$ for small values of the field $h$.  
The zero order term in the expansion vanishes and we
obtain a leading linear term:
\beq 
\rho_a(h)=\frac{uh}{1+u-ug+u\epsilon}.
\label{rhoa}
\eeq
This result has to be consistent with 
the global conservation law, which states that the average
input energy flux $J_{in}$ must balance the dissipated flux $J_{out}$.
In the stationary state the conservation law can be written as
\beq 
J_{in}=hL^d=J_{out}=\epsilon\rho_aL^d.
\label{cons}
\eeq
By comparing Eq.~\ref{rhoa} with Eq.~\ref{cons} we obtain
that $u=1/(g-1)$. In the limit $h\to 0$ the densities 
are therefore given by
\beq
\rho_a=\frac{h}{\epsilon},~~~~~\rho_c=\frac{1}{g}+{\cal O}(h),
~~~~\rho_s=\frac{g-1}{g}+{\cal O}(h).
\eeq
An estimate of $g$ can be obtained using a random neighbor
approximation, which yields $g=2d$ for the BTW model \cite{bak2}
or $g=2$ for the two level models \cite{manna}.
Noticeably, in the latter case $u=1$, i.e. all stable sites are 
sub-critical, as it is expected for a two level model.

We now discuss the critical behavior of these systems.
The balance between conservation laws and dissipation is essential
for the critical behavior of the model, as it has been also pointed out
in \cite{sobp}. 
The model is critical just in the 
double limit $h,\epsilon \to 0, h/\epsilon\to 0$, similarly to the forest-fire 
model\cite{dro}.
In analogy with non equilibrium phenomena \cite{torre,jmendes}, 
the one particle density of active sites is the {\em order parameter}
and goes to zero at the critical point. 
We can then distinguish several different regimes as
a function of the parameters. The system has no stationary state for
$h>\epsilon$, since $\rho_a$ would have to be greater than one to 
satisfy Eq.~(\ref{cons}). The model is supercritical for
$h>0$ and $\epsilon>h$, while for $h\to 0$ and $\epsilon>0$ 
it is subcritical and the dynamics
displays avalanches. The phase diagram  is somehow similar to
that of usual second order phase transitions, if we replace 
$h$ by the magnetic field
and $\epsilon$ by the reduced temperature. 

In the supercritical regime the order parameter is linear in $h$ 
\beq
\rho_a\sim h^{1/\delta};~~~~~~~~\delta=1.
\eeq
This is analogous to the MF results obtained for contact processes and 
other non equilibrium CA \cite{torre,dick,jmendes}, but it is 
in contrast with previous MF approaches for sandpile models
\cite{bakmf,stellamf}, which yielded $\delta=2$ \cite{notemf}.
This incorrect result is due to an inconsistency
present in those studies. The scaling is expressed in terms of 
the average energy $\theta \equiv \sum_i \rho_i z_i$ which is
treated as an independent control parameter. As we have just 
shown, $\theta$ and $h$ {\em are not independent}. Moreover,
$\theta$ can not be considered as the control 
parameter even for $h=0$, since 
it does not determine completely the state of the system: 
the same value of $\theta$ describes several states corresponding to
different values of densities $\rho_i$. This is a typical
property of CA with multiple absorbing states \cite{jmendes}.
In analogy with non equilibrium CA it is possible to define several other 
exponents characterizing the supercritical regime  for this class of models.
The complete set of exponents and their scaling relations will be 
reported elsewhere \cite{vz}.

In the subcritical regime , 
the behavior of the system is dominated by the dissipation.
This can be seen by studying the susceptibility  
\beq
\chi\equiv\frac{\partial \rho_a(h)}{\partial h}=\frac{1}{\epsilon},
\eeq
which diverges for $\epsilon=0$. 
The system is in a subcritical state 
for any value of $\epsilon$ different from zero.
The critical behavior is thus characterized by the scaling laws 
$\chi\sim \epsilon^{-\gamma}$ and $\xi\sim \epsilon^{-\nu}$,
where $\xi$ is the characteristic length. 

We can use these exponents to characterize the 
conservative sandpile model, since
our MF analysis treats both border and bulk dissipation.
In conservative systems, when the size is increased  
the effective dissipation
decreases as $\epsilon\sim L^{-\mu}$, 
since bulk processes dominate over boundary dissipation.
At the same time, the characteristic length of the avalanches should go
like $\xi\sim L$ to ensure dissipation of energy. 
This implies the scaling relation $\nu\mu=1$, and that 
$\chi\sim L^{\mu\gamma}$. 
It is also possible to show\cite{vz},
that the susceptibility scales as the average avalanche size,  
and in two dimensions it has been found that 
$\langle s\rangle\sim L^2$ exactly for $L\to\infty$ \cite{dhar}. 
The same result holds also
in MF theory if we assume that the dynamics is diffusion like
\cite{vz,zhang,co}.

Combining all the above results we obtain a first set of MF exponents
\beq
\gamma=1, ~~~~\mu=2, ~~~~\nu=1/2.
\eeq
It is worth to remark that it is not possible to define the equivalent 
of an exponent $\beta$, because for $h=0$ the order parameter is always zero.
We emphasizes again that $h$ and $\epsilon$ are both control parameter 
responsible for different regimes of the model.   

\begin{figure}[htb]
\narrowtext
\centerline{
        \epsfxsize=7.0cm
        \epsfbox{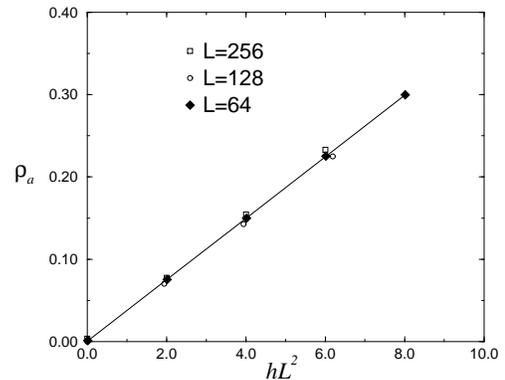}
        }
\caption{The density of active site in the BTW model
with border dissipation as a function of the driving rate $h$.}
\label{fig:1}
\end{figure}

\begin{figure}[htb]
\narrowtext
\centerline{
        \epsfxsize=7.0cm
        \epsfbox{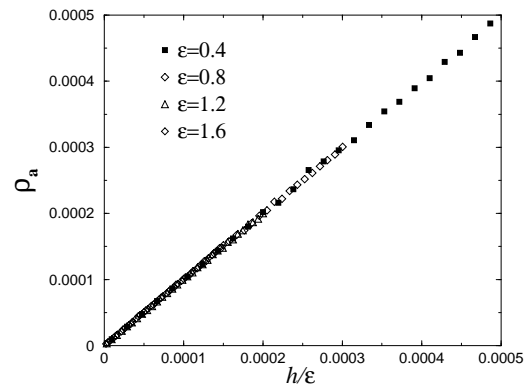}
        }
\caption{The density of active sites for the BTW model
with bulk dissipation $\epsilon$ and periodic boundary
conditions as a function of $h/\epsilon$ for L=64.}
\label{fig:2}
\end{figure}

We have derived these exponents using only conservation
laws, therefore we expect they should describe also low
dimensional sandpile models. 
We simulate numerically the BTW model with finite driving rate 
$h$ and boundary dissipation. We see in Fig~\ref{fig:1}
that the density of critical sites goes to zero linearly with
$h$ ($\delta=1$) with a slope that increases with the system size 
as $L^{2}$. This is in agreement with the MF theory which
predicts that the susceptibility scales as $L^{\mu\gamma}$,
with $\mu\gamma=2$.
To observe more clearly the scaling with dissipation 
of the sandpile model we study the BTW model 
with {\em periodic} boundary
conditions and fixed dissipation $\epsilon$ \cite{defdiss}. 
In Fig.~\ref{fig:2} we plot the control 
parameter as a function of $h/\epsilon$.
The scaling predicted by the MF theory ($\gamma=1$) is verified with 
remarkable accuracy, and we note that
finite size corrections are not noticeable, in contrast with
the case of boundary dissipation\cite{grasma}.
Finally, the exponent $\nu =1/2$ has been measured already in
a two dimensional dissipative sandpile model \cite{diss}.

The dynamics in the subcritical regime  takes place
in the form of avalanches. The exponents describing
avalanche distributions depends on the dimension
and in general will not agree with the MF results.
A complete characterization of MF avalanche
scaling has been obtained by using the theory
of branching processes\cite{sobp}.
Here, we reproduce these results in an independent way. 
Following
Grassberger and de la Torre \cite{torre} we consider 
the probability that a small perturbation  activate 
$s$ sites (an avalanche in the SOC terminology)
\beq
P(s,\epsilon)=s^{-\tau}{\cal G}(s/s_c(\epsilon)),
\eeq
where $s_c\sim\epsilon^{-1/\sigma}$ is the cutoff in the avalanche size. 
The perturbation decays in the  stationary subcritical state as 
\beq
\rho_a(t) \sim t^{\eta}{\cal F}(t/t_c).
\eeq
Here $t_c$ denotes the characteristic time which scales as
$t_c \sim \epsilon^{-\Delta}$. We can obtain these exponents 
by solving Eq.~(\ref{ssmf}) for a small perturbation
around the subcritical stationary state ($h=0$)
\beq
\frac{\partial \rho_a}{\partial t}= -\rho_a+(g-\epsilon)\rho_a\rho_c,
\eeq
where, as a first order approximation, we can replace $\rho_c=1/g$. 
We obtain in this way $\rho_a(t)\sim\exp(-\epsilon t/g)$
which implies $\eta=0$ and $\Delta=1$.    
Introducing the scaling laws  $s_c\sim\xi^{D}$ and $t_c\sim\xi^z$,
it is possible to derive \cite{vz} another set of scaling relations
\beq
\gamma=\frac{(2-\tau)}{\sigma},~~~D=\frac{1}{\nu\sigma},~~~
z\nu=\Delta,~~~\frac{(\tau-1)}{\nu\sigma}=z ;
\eeq
from which we get the second set of MF critical exponents
\beq
z=2,~~~~D=4,~~~~~\tau=3/2,~~~~~\sigma=1/2;
\eeq
in agreement with the theory of branching processes \cite{sobp}.
It is worth to remark that the numerical value of these exponents
is the same as in other MF approach \cite{bakmf,stellamf}, but their 
significance is completely different being defined  with respect to 
a different scaling field.

We have obtained a complete characterization of the critical
properties of the sandpile model. The critical state arises
due to the fine tuning of the driving rate and the dissipation.
This condition is enforced implicitly in the BTW model by imposing
time scale separation and dissipation only
through the boundaries, which makes $h$ and
$\epsilon$ equal to zero in the thermodynamic limit.
In this formalism SOC appears as a special case
of non equilibrium critical phenomena, with the only peculiarity
that the critical parameters are zero. The same MF analysis
applied to the forest-fire model leads to similar
conclusions \cite{vz,mfff}. Contrary to sandpile models
in the forest-fire model no conservation laws are present 
and MF exponents are not correct in low dimensions.
It would be desirable to treat extremal models \cite{extrem} 
in a similar way, emphasizing the role of the driving rate.
We hope this will clarify the precise significance of SOC
in the framework of non equilibrium critical phenomena.

The Center for Polymer Studies is supported by NSF.


\end{multicols}

\begin{thebibliography}{99}
\bibitem{disboun}
        The boundary dissipations  is going to zero for infinite 
        systems, however this limit can not be performed before imposing 
      	$h\to 0$, because of global conservation.
\bibitem{notemf}
        In nonequilibrium CA this exponent is usually denoted 
        by $\beta$. We keep the notation of 
        Ref.~\cite{bakmf,stellamf} to emphasize the differences
        in the results we obtain.
\bibitem{defdiss}
        To allow for bulk dissipation, we introduce 
        a probability that the energy involved 
        in a relaxation event is lost.
\end{thebibliography}
\end{document}